%% file: ms.tex
\documentclass[preprint2]{aastex}

\slugcomment{Submitted to ApJ}

\shorttitle{Runaway Massive Binaries}
\shortauthors{McSwain et al.}

\begin{document}

\title{Runaway Massive Binaries and Cluster Ejection Scenarios}

\author{M.\ Virginia McSwain\altaffilmark{1}}
\affil{Department of Astronomy, Yale
University, P.O. Box 208101, New Haven, CT 06520-8101}
\email{mcswain@astro.yale.edu}

\author{Scott M.\ Ransom}
\affil{National Radio Astronomy Observatory, 520 Edgemont Road, 
Charlottesville, VA 22903}
\email{sransom@nrao.edu}

\author{Tabetha S.\ Boyajian, Erika D.\ Grundstrom}
\affil{Department of Physics and Astronomy, Georgia State University,
P.O. Box 4106, Atlanta, GA 30302-4106}
\email{tabetha@chara.gsu.edu, erika@chara.gsu.edu}

\author{Mallory S.\ E.\ Roberts}
\affil{Eureka Scientific, Inc., 2452 Delmer Street Suite 100, Oakland, CA 
94602-3017}
\email{malloryr@gmail.com}

\altaffiltext{1}{NSF Astronomy and Astrophysics Postdoctoral Fellow}


\def\kms    {\ifmmode{{\rm km~s}^{-1}}\else{km~s$^{-1}$}\fi}
\def\Mdot   {\ifmmode {\dot M} \else $\dot M$\fi}
\def\Mspy   {\ifmmode {M_{\odot} {\rm yr}^{-1}} \else 
$M_{\odot}$~yr$^{-1}$\fi}
\def\Msun   {$M_{\odot}$}
\def\mum     {\ifmmode{\mu{\rm m}}\else{$\mu{\rm m}$}\fi}
\def\Rstar  {$R_{\star}$}


\begin{abstract}

The production of runaway massive binaries offers key insights into the
evolution of close binary stars and open clusters.  The stars HD 14633 and
HD 15137 are rare examples of such runaway systems, and in this work we
investigate the mechanism by which they were ejected from their parent
open cluster, NGC 654.  We discuss observational characteristics that can
be used to distinguish supernova ejected systems from those ejected by
dynamical interactions, and we present the results of a new radio pulsar
search of these systems as well as estimates of their predicted X-ray flux
assuming that each binary contains a compact object.  Since neither
pulsars nor X-ray emission are observed in these systems, we cannot
conclude that these binaries contain compact companions.  We also consider
whether they may have been ejected by dynamical interactions in the dense
environment where they formed, and our simulations of four-body
interactions suggest that a dynamical origin is possible but unlikely.  
We recommend further X-ray observations that will conclusively identify
whether HD 14633 or HD 15137 contain neutron stars.

\end{abstract}



\section{Introduction}

While most O- and B-type stars are believed to form in open clusters and
stellar associations, some are observed at high galactic latitudes and
with large peculiar space velocities.  The estimated fraction of O stars
that are runaways ranges from 7\% \citep*{conti1977} to 49\%
\citep{stone1979}.  These stars were likely ejected from the clusters of
their birth, and there are two accepted mechanisms to explain the origin
of their runaway velocities.  Close multi-body interactions in a dense
cluster environment may cause one or more stars to be scattered out of the
region \citep*{poveda1967}.  On the other hand, a supernova explosion
within a close binary may eject the secondary star due to the conservation
of momentum of mass lost in the explosion, including supernova kicks
\citep{zwicky1957, blaauw1961, sutantyo1978}.

The production of runaway O- and B-type binaries is expected to be rare,
but these systems can offer key insights into the evolution of close
binary stars and open clusters.  Distinguishing between the dynamical
ejection and binary supernova scenarios for an isolated runaway star can be
nearly impossible unless the star contains CNO-processed gas received from
a past episode of mass transfer \citep*{hoogerwerf2001}.  However, the
ejection mechanism responsible for producing runaway binaries can be
distinguished more easily from a variety of observable properties.

\citet{leonard1990} use $N$-body simulations of open clusters to show that
a binary frequency of about 10\% is expected from dynamical ejections.  
Most of their simulated binaries have eccentric orbits, $0.4 < e < 0.7$,
and high mass ratios, $q = M_2/M_1 > 0.5$.  While \citet{leonard1990} do
not discuss the resulting orbital periods in their simulations, we do not
expect long period ($P > 1000$ days) binaries to form via dynamical
interactions since their low binding energies would likely cause them to 
be ionized.  On the other hand, \citet{portegieszwart2000} predicts a
higher binary fraction of $20-40$\% among runaways which are ejected by
supernovae in close binaries.  The resulting neutron star may remain bound
to the secondary if not enough mass is lost during the explosion or if the
explosion is not symmetric (producing a kick velocity).  In such cases,
the present-day orbital period and eccentricity depend on the mass ejected
during the supernova and the kick velocity \citep*{nelemans1999,
brandt1995}, but the resulting period distribution is expected to be
$3-1000$ days \citep{portegieszwart2000}.

Distinguishing between the dynamical or supernova ejection mechanisms
among runaway spectroscopic binaries can be a difficult task since both
are expected to produce eccentric, relatively short period binaries.  
Only dynamical interactions are expected to produce double-lined
spectroscopic binaries (SB2s), while a single-lined system (SB1) may be
formed either way. Obtaining very high signal-to-noise (S/N) spectra or
performing doppler tomography can, in some cases, identify optical
companions in such systems.  However, it can be nearly impossible to 
detect a cool, low mass, optical companion with an O- or B-type primary.

Neutron star companions can be difficult to identify as well.  A few
runaway O stars are predicted to be associated with pulsars
\citep{portegieszwart2000}, but so far none have been found
\citep*{philp1996, sayer1996}.  X-ray emission may also be used to
pinpoint neutron star companions since a neutron star may accrete mass
from its companion via a Roche lobe overflow stream or by Bondi-Hoyle
accretion from the stellar wind of the luminous primary \citep{bondi1944,
davidson1973, kaper1998}.  The systems experiencing Roche lobe overflow
usually have large X-ray luminosities and striking optical emission lines.  
On the other hand, weaker X-ray emission will result if stellar winds are
the accretion source. According to the wind accretion model, the observed
X-ray luminosity depends on the system separation, the relative wind and
companion velocities, the stellar mass-loss rate, and the mass of the
accretor \citep*{lamers1976}.  If the stellar wind is too weak, or the
system separation too large, the X-ray luminosity would diminish below the
detection limit of today's X-ray surveys, hence a ``quiet'' massive X-ray
binary (MXRB).

We have recently completed a radial velocity study of 12 field and runaway
O- and B-type stars to identify spectroscopic binaries with high space
velocities \citep{mcswain2007}.  In that work, we improved the orbital
elements of five SB1s, but among these only two are definitely runaways:
HD 14633 and HD 15137. HD 14633 is classified as an ON8.5 V star, and HD
15137 is somewhat more evolved as an O9.5 III(n) star.  
\citet{boyajian2005} found that these runaway SB1 systems were likely
ejected in separate events from the open cluster NGC 654, but the travel
times since their ejection (14 Myr for HD 14633 and 10 Myr for HD 15137)
present an unusual paradox since they are longer than the expected
lifetimes of O-type stars.  They attempted doppler tomographic separation
for both systems, but this search for an optical companion was
unsuccessful.  HD 14633 is a nitrogen-strong star \citep{walborn1972}, and
Boyajian et al.\ argue that its nitrogen enrichment and fast 
rotation resulted from a mass
transfer episode prior to a supernova in the close binary system.  They
propose that both stars are quiet MXRB candidates that harbor neutron star
companions but are not known X-ray emitters.

To investigate this claim, we present here the results of a pulsar search
in these two systems.  We also compare the observed upper limits of their
X-ray fluxes to model predictions of X-rays generated by the accretion of
stellar winds onto a compact companion.  Since neither pulsars nor X-ray
emission are detected, we are not certain that either system contains the
product of a supernova.  Therefore we also investigate the possibility
that dynamical interactions may have been responsible for their ejection
from NGC 654.  


\section{Constraints on the Secondary Masses}

\begin{figure}
\includegraphics[angle=0,scale=.5]{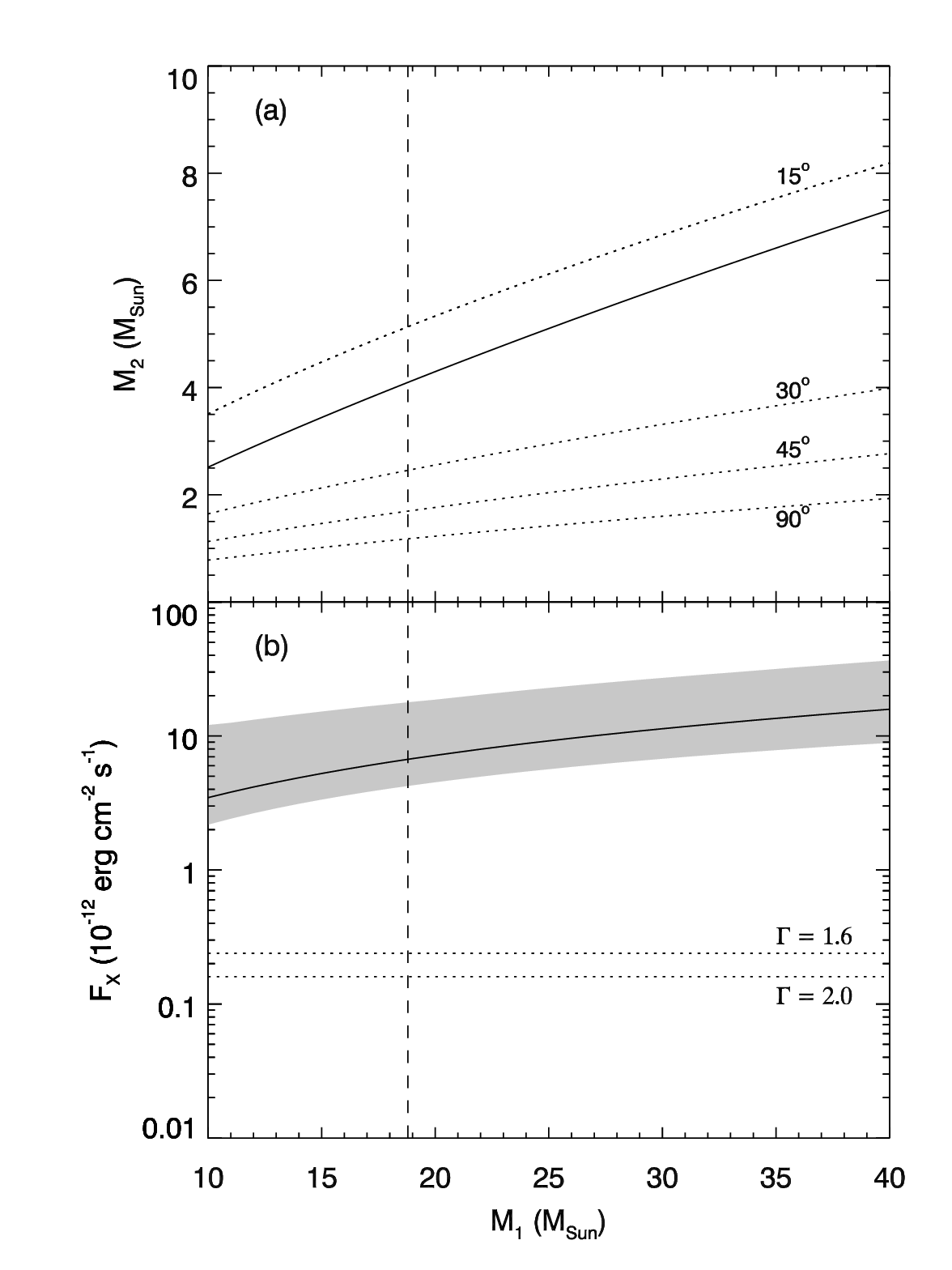}
\caption{
\label{hd14633}
(a) The mass diagram of HD 14633 is plotted for a range of
inclination angles (\textit{dotted lines}).  The most probable values of
$M_2$ from \citet{mazeh1992} (\textit{solid line}) and the 
expected $M_1$ from \citet{martins2005} (\textit{vertical dashed line}) 
are also shown.  
(b) Predicted time-averaged X-ray flux (\textit{solid line}) and the 
predicted range in flux over the eccentric orbit 
(\textit{gray}) are shown for the lowest possible $M_2$.  
The observed upper limits for $F_X$ are shown for comparison 
(\textit{dotted lines}).
}
\end{figure}

\begin{figure}
\includegraphics[angle=0,scale=0.5]{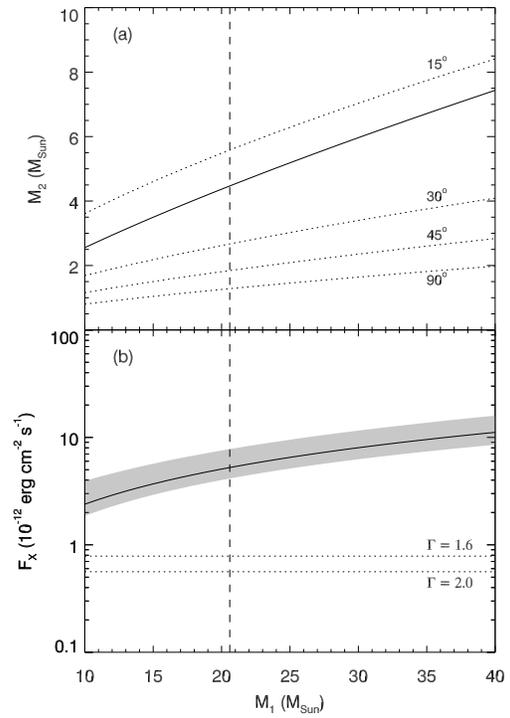}
\caption{   
\label{hd15137}
The mass diagram and predicted X-ray flux are plotted for HD 15137 in the 
same format as Figure \ref{hd14633}.
}
\end{figure}

We recently presented updated orbital elements and mass functions, $f(m)$,
for the SB1 systems HD 14633 and HD 15137 \citep{mcswain2007}, and these
parameters are summarized in Table \ref{paper1}.  While $f(m)$ does not
strictly define the component masses, it does allow constraints on the
primary star's mass, $M_1$, the companion mass, $M_2$, and the system
inclination, $i$.  We plot the relationships between these properties in
Figures \ref{hd14633} and \ref{hd15137} using a grid of values
for $M_1$.  In addition, the vertical dashed line in both figures
indicates the expected mass of each star based on the calibration of
\citet*{martins2005}.  Both binaries have very low $f(m)$ that suggest low
mass companions ($1-3 M_\odot$).  Without directly detecting the
companions, we must constrain their masses using another method.

\placetable{paper1}
\placefigure{hd14633}
\placefigure{hd15137}

We used the statistical method of \citet{mazeh1992} to determine the most
probable $q$, and this relationship is plotted as a solid line in the top
panels of Figures \ref{hd14633} and \ref{hd15137}.  Note that
their technique results in a somewhat lower probable value for $i$ than
the expectation value $\langle i \rangle \sim 60^\circ$ since it accounts
for observational selection effects.  HD 14633 and HD 15137 are
particularly interesting since they probably have $q < 0.3$, and such low
$q$ are unknown among O star binaries (excluding X-ray binaries).  Their
allowed masses do not exclude neutron star companions, and below we
investigate the possibility of radio pulsars or a detectable X-ray
signature of a compact companion.


\section{Pulsar Search Results}

HD 14633 was targeted in a previous search for pulsars in runaway OB stars
by \citet{philp1996}.  Although they did not find any pulsars, we chose to
perform a new, more sensitive search to investigate the companions of both
runaway binaries.  \citet{portegieszwart2000} shows that detection of
pulsars in O star binaries is hindered by the absorption of the radio
emission by the stellar winds.  This absorption will be greater at
periastron than apastron since the winds are densest close to the O star,
and our new ephemerides for these binaries \citep{mcswain2007} allowed us
to schedule the observations as close as possible to the maximum
separation.  (If these systems have high $i$, true if a neutron star is
present, the times of inferior conjunctions may provide the lowest column
depth instead.  These times correspond to the orbital phases 0.83 for HD
14633 and 0.70 for HD 15137.  Although our observations were not timed to
occur at inferior conjunction, they are well separated from superior
conjunction when the neutron star could possibly be eclipsed.)  We also
chose to use higher frequencies than the previous investigations because
they are more likely to detect radio pulses dispersed by the stellar
winds.

We observed HD 14633 and HD 15137 using the National Radio Astronomy
Observatory's 100-m Green Bank Telescope (GBT)  and the Pulsar Spigot
back-end \citep{kaplan2005}.  We used the Spigot in mode 2 with the S-band
receiver to obtain 600 MHz (1650-2250 MHz) of bandwidth, which the Spigot
summed and synthesized into 768 0.78125-MHz frequency channels every 81.92
$\mu$s.  The Spigot in mode 42, with the 820 MHz receiver, provided 50 MHz
of bandwidth (795-845 MHz), summed into 1024 0.0488-MHz frequency channels
with the same time resolution.  Each target was observed once with each
receiver for 6300 s.  For each observation, the time at midexposure and
the corresponding orbital phase are listed in Table \ref{gbt}.  Although
we requested observations at the times of apastron (orbital phase 0.5),
telescope scheduling constraints limited us to slightly earlier times
during each orbit.

\placetable{gbt}

The data were reduced using the PRESTO software package
\citep{ransom2001}.  We first searched the raw data for radio frequency
interference in both the time and frequency domains and applied
interference masks to remove these sources.  Each of our targets is
expected to have a dispersion measure (DM) $\sim 50-75$ pc~cm$^{-3}$ using
the NE2001 model \citep{cordes2002} with the distances in Table
\ref{paper1} \citep{mcswain2007}, so we searched the observations by
dedispersing the raw data into separate time series with DMs ranging from
0 to 120 pc~cm$^{-3}$ and spaced by 0.5 pc~cm$^{-3}$.  We then
Fourier-transformed each time series and searched them by using
Fourier-domain acceleration search techniques in order to maintain
sensitivity to both pulsars in binary orbits as well as isolated pulsars.  
For the acceleration search, this technique involves a linear
approximation of the binary's orbital acceleration (valid since the
observations span a small fraction of the orbit) to look for a derivative
in pulsar frequency.  The flux density sensitivity, $S_{\rm lim}$, and
luminosity, $L_{\rm lim}$, limits of each observation are listed in Table
\ref{gbt}.  The typical pulsar has a luminosity $L \gtrsim
0.1$~mJy~kpc$^2$ at 1400 MHz \citep{lorimer2006}, and our $L_{\rm lim}$ do
go fainter than this value for the S-band observations.


While we did not detect any pulsars in these systems, we cannot rule out
their existence since a pulsar's beam may not cross our line of sight.  
\citet{tauris1998} find that the observed beaming fraction depends on the
rotational period, $P$, since the beam radius varies as $P^{-1/2}$.  $P$,
in turn, depends on the pulsar's age, $t$, since magnetic dipole radiation
causes the pulsar to spin down over time.  Without detecting a pulsar we
cannot measure $P$ directly, but if the runaways were ejected by
supernovae, their dynamical ages correspond to the ages of the resulting
neutron star.  HD 14633 was ejected from the open cluster NGC 654 about 14
Myr ago, and HD 15137 left the same cluster about 10 Myr ago
\citep{boyajian2005}.  If we assume that any pulsar was born spinning much
faster than its current rate, we can use
\begin{equation} 
P \approx \sqrt{\frac{8 \pi^2 R^6}{3 c^3 I} \: B^2 \: t},
\end{equation} 
where $R$ is the neutron star radius, $c$ is the speed of light, $I$ is
the moment of inertia, $B$ is the strength of the magnetic field, and $t$
is the pulsar age in seconds \citep{tauris1998}.  Using typical neutron
star values of $R = 10$~km, $I = 10^{45}$~g~cm$^2$, and $B = 10^{12}$~G
\citep{lipunov1992}, we estimate $P \approx 1$~s for HD 14633 and HD
15137.  \citet{lorimer2006} find a beaming fraction of $\sim 20$\% for 
such pulsars, making it unlikely that we will detect the beam.

Clearly, our GBT observations are not sufficient to characterize the
companions in either system.  Below we investigate their other properties
to determine if they should display an observable X-ray luminosity.


\section{Predicted X-ray Luminosities}

\subsection{Wind Accretion Model}

HD 14633 and HD 15137 are SB1s with high eccentricity, but neither
are known X-ray sources.  If they do contain compact companions, should
these stars have a detectable X-ray emission?  We predict the X-ray flux of 
each quiet MXRB candidate using the wind accretion model of 
\citet{lamers1976}, modified for eccentric orbits.
This method uses the Bondi-Hoyle accretion rate 
\begin{equation}
S_a = \frac{\pi R_a^2 \Mdot}{4 \pi a^2}
\end{equation}
\citep{lamers1976}, which depends on the stellar mass loss rate, \Mdot, the 
system's major axis, $a$, and the accretion radius of the companion,
\begin{equation}
R_a = \frac{2 G M_2}{V_{rel}^2}.
\end{equation}
Here, $M_2$ is the mass of the accretor and $V_{rel}$ is the relative 
velocity of the stellar wind and the companion's orbit, $V_{wind}$ and 
$V_{orb}$ respectively.  $V_{orb}$ is determined from $a$ and the orbital 
eccentricity, $e$.  The predicted X-ray luminosity is 
\begin{equation}
L_X \propto \zeta S_a M_2
\end{equation}
\citep{lamers1976}.  The efficiency, $\zeta$, of converting accreting 
matter into X-ray luminosity is approximately 0.1 for neutron stars and 
black holes \citep{lamers1976}.  The observed X-ray flux will also depend 
on the system distance, $d$, and the \ion{H}{1} column density, nH.  While 
the Lamers et al.\ model is useful for estimating the integrated X-ray 
flux produced by wind accretion, we emphasize that it is an approximation 
that does not include any spectral information about the emission.  
Therefore the predicted X-ray emission should be viewed with some caution.

We could not measure the mass loss rate, \Mdot, from our optical spectra
\citep{mcswain2007} since we did not observe the H$\alpha$ line.  Instead
we obtained published values of the observed \Mdot~for each star.  
\citet{chlebowski1991} and \citet{howarth1989} found values for HD 14633
of $\log$ \Mdot = $-6.81$~ $M_\odot$~yr$^{-1}$~ and $-6.9$~
$M_\odot$~yr$^{-1}$, respectively.  We adopt the lower value for our model
of the wind accretion.  \citet{howarth1989} also measured $\log$ \Mdot =
$-6.5$ ~$M_\odot$~yr$^{-1}$~ for HD 15137.

We measured their terminal wind velocities, $V_\infty$, using
flux-calibrated UV spectra obtained from the \textit{International
Ultraviolet Explorer (IUE)} archives. \citet*{prinja1990} describe how to
use either the time-variable narrow absorption components of UV lines or
the more steady-state, saturated line edge to measure $V_\infty$, and we
chose this latter technique due to the low S/N of the \textit{IUE}
spectra.  For HD 14633, we co-added the available high dispersion, Short
Wavelength Prime Camera (SWP) spectra and looked for saturation in the
\ion{N}{5} $\lambda$1238.821, \ion{Si}{4} $\lambda$1393.76, and \ion{C}{4}
$\lambda$1548.202 lines.  $V_\infty$ corresponds to the blue edge of the
saturated line profile.  HD 15137 has only one high dispersion SWP
spectrum available, so we binned this by 5 pixels to improve the S/N,
reducing the precision of our measurement.  The resulting values of
$V_\infty$ are listed in Table \ref{termvel}, and they agree well with the
values found by \citet{howarth1997}.  Knowledge of $V_\infty$ enables us
to characterize the stellar wind in the vicinity of the star according to
the wind velocity law
\begin{equation} 
V(r) = V_{\infty} \left(1-\frac{R_\star}{r}\right)^\beta
\end{equation} 
with $\beta$ = 0.8 for weak stellar winds \citep{howarth1989, puls1996}.

\placetable{termvel}

The remaining input parameters for the wind accretion model come from our
orbital solutions, SED fits, and stellar distances \citep{mcswain2007}.  
We used a grid of values for $M_1$ with the minimum $M_2$ allowed by the
mass functions ($i = 90^\circ$).  For these eccentric binary orbits, we
expect the accretion rate to be greatest just after periastron, where the
wind density is higher and its velocity slower relative to the companion.  
Therefore we calculated the time-averaged X-ray flux, $F_{X}$, as well as
the expected maximum and minimum flux during the orbit as a function of
the primary mass.  The results are illustrated in the bottom panels of
Figures \ref{hd14633} and \ref{hd15137}.  We note that a higher
$M_2$ increases the predicted X-ray flux since the gravitational force
accreting the wind is larger.

Normal O stars are expected to have an intrinsic X-ray flux due to
shocks in their stellar winds (\citealt{sana2006} and references therein).  
The winds are driven by the absorption and reemission of UV photons in the
spectral lines, and small-scale instabilities in the wind will quickly
grow into shocks \citep{lamers1999}.  Magnetic confinement of the stellar
winds and colliding winds in close, massive binaries may also contribute
to the X-ray emission (\citealt{sana2006} and references therein).  
\citet{sana2006} found that, with the exception of two colliding wind
binaries, all of the binary and single O-type stars in the open cluster
NGC 6231 display a tight correlation between their bolometric
luminosities, $L_{bol}$, and their X-ray luminosities, $L_X$, in the
$0.5-10.0$ keV band.  Thus the intrinsic X-ray emission of an O star can
be predicted from the relation, 
\begin{equation}
\log L_X - \log L_{bol} = -6.912 \pm 0.153
\end{equation}
\citep{sana2006}. Using $L_{bol}$ from the calibration of 
\citet{martins2005}, we find that HD 14633 and HD 15137 each have an 
intrinsic $L_X$ of $2\times 10^{31}$ to $3 \times 10^{31}$ erg~s$^{-1}$.  
With the adopted distances given in Table \ref{paper1} 
\citep{mcswain2007}, this corresponds to an unabsorbed flux between $3 
\times 10^{-14}$ and $7 \times 10^{-14}$ erg~cm$^{-2}$~s$^{-1}$, a 
negligible contribution to the predicted flux from stellar wind accretion.

\subsection{Observed Upper Limits}

Since these stars have not been detected as X-ray sources, we estimate an
upper limit for their unabsorbed X-ray fluxes using other detected sources
from the \textit{ROSAT} All-Sky Survey and pointed \textit{ROSAT}
observations within a $30\arcmin$ radius of each star \citep*{white2000,
voges2000}.  We expect that wind accretion will generate X-rays having a
power law spectrum typical of MXRBs, and we used the photon index $\Gamma =
2.0$ from the microquasar LS 5039 (another wind-accreting MXRB;
\citealt{boschramon2005}).  MXRBs in a low/hard emission state generally
have $\Gamma \sim 1.6$ \citep{belloni2001}, which we also consider.  The
Galactic \ion{H}{1} column density, nH, for each target was calculated
using NASA's HEASARC tool\footnote{The HEASARC nH calculator is available
online at http://heasarc.nasa.gov/cgi-bin/Tools/w3nh/w3nh.pl.}.  Finally,
we used the HEASARC WebPIMMS calculator\footnote{The WebPIMMS calculator is
available online at http://heasarc.nasa.gov/Tools/w3pimms.html.} to convert
the upper limit of the X-ray count rate to the expected unabsorbed flux in
the $0.5-10$ keV range.  These details and the resulting upper limits for
$F_X$ are listed in Table \ref{xraylimits}.  We compare these upper limits
to the predicted $F_X$ in the bottom panels of Figures \ref{hd14633}
and \ref{hd15137}.  In each case, the observed upper limit is higher
than the intrinsic X-ray flux from the O-star, consistent with the
non-detection of each star.

The wind accretion model predicts X-ray fluxes of HD 14633 and HD 15137
that are at least an order of magnitude greater than the observed upper
limits for the most likely O star masses.  However, it is worth noting
that the wind-accreting microquasar LS 5039 exhibits order-of-magnitude
variations in both its X-ray flux and mass-loss rates
\citep{boschramon2005, mcswain2004}, and other MXRBs may have flux
variations of order 100 \citep{belloni2001}. If the \textit{ROSAT}
observations occured during times of low stellar mass loss, it is possible
that any X-ray emission from a neutron star might be below the detection
limit.  O star winds may also attenuate any X-ray emission in low energy
bands, and \citet{sana2004} find order-of-magnitude variations in the
X-ray emission from the colliding wind binary HD 152248 due to variations
in the absorption with orbital motion.  If HD 14633 and HD 15137 are high
inclination systems, they may experience similar variations.

\subsection{Quiescent Neutron Stars}

Could these binaries contain neutron stars in a quiescent state?  The
conditions for wind accretion onto a neutron star depend strongly upon the
spin rate and the magnetic field, and these are discussed in detail by
\citet{lipunov1992}.  A young neutron star will not accrete significant
amounts of material because its fast rotation and large magnetic field
sweep material out of the system at a distance larger than the accretion
capture radius, $R_a$.  Systems in this ejector regime are often observed
as pulsars, and they are expected to remain in this state for $10^5-10^6$
years.  The spinning magnetic field will dissipate rotational energy,
causing the neutron star to spin down with time.  The neutron star passes
into the propellor regime, where the magnetospheric radius, $R_{mag}$, is
smaller than $R_a$ but remains larger than the corotation radius, $R_c$.  
The rapid rotation continues to inhibit accretion onto the neutron star.  
Eventually the compact star spins down enough to allow $R_c$ to expand
outward beyond $R_{mag}$, trapping material in the vicinity of the neutron
star and allowing accretion to begin.

We can investigate whether the proposed neutron stars should be in the 
accretion regime using these characteristic distances given by 
\citet{lipunov1992}.  Assuming a nominal $M_2 = 1.4~M_\odot$ for the 
neutron star and using the relative wind/orbital velocity $V_{rel} \sim 
1500$~\kms~ for both systems, their accretion capture radius is
\begin{equation}
R_a = \frac{2 G M_2}{V_{rel}^2} \sim 2*10^{10} \; \rm cm.
\end{equation}
Using the rotational frequency $\omega \approx 1$ s$^{-1}$ (discussed in 
Section 3), we can estimate the corotation radius as
\begin{equation}
R_c = \left( \frac{G M_2}{\omega^2} \right)^{1/3} \sim 6*10^8 \; 
\rm cm.
\end{equation}
The magnetospheric radius, $R_{mag}$, can be determined by equating the ram 
pressure of the accreting gas to the magnetic field pressure.  While we 
cannot measure the magnetic moment, $\mu$, the observed range among other 
neutron stars is $10^{28}$ to $10^{31.5}$ G~cm$^{3}$, with the most typical  
value being $10^{30}$ G~cm$^{3}$ \citep{lipunov1992}.  Thus we find 
\begin{equation}
R_{mag} = \left( \frac{\mu^2}{2 S_a \sqrt{2 G M_2}} \right)^{2/7} 
\sim 3*10^9 \; \rm cm
\end{equation}
for both binaries.  Unrestrained accretion onto the neutron star requires 
that $R_{mag} < R_a$ and $R_{mag} < R_c$, which may be possible in these 
systems given the large uncertainties in $R_c$ and $R_{mag}$. 

HD 14633 and HD 15137 could contain neutron stars in a low accretion
state, and new X-ray observations should conclusively show whether these
systems harbor compact companions.  Even quiescent neutron stars are
expected to produce an X-ray luminosity of $10^{30}$ to $10^{34}$
erg~s$^{-1}$ with a thermal or nonthermal spectrum \citep{lipunov1992},
somewhat comparable in magnitude to the expected intrinsic luminosity of
an O star but presumably distinguishable by the spectral energy
distribution.  Therefore any detection of X-rays with some spectral
resolution should be able to distinguish between the soft spectrum of
stellar wind emission, the presence of a quiescent neutron star, or
emission produced by wind accretion with a power law spectrum.


\section{The Dynamical Interaction Scenario}

Both HD 14633 and HD 15137 contain a secondary star with a minimum mass
$M_2 = 1-2~M_\odot$, so a neutron star companion seems plausible.  
However, we do not detect pulsars in either system, and neither HD 14633
nor HD 15137 exhibits evidence of X-rays produced by accretion onto a
compact companion. Although we have discussed several reasons to account
for the lack of radio or X-ray detections, we must explore the possibility
that neutron stars are not present in these systems.  We consider here
whether these runaway SB1s may have been ejected from NGC 654 by close
dynamical encounters in the open cluster.

In our earlier study of these runaway stars \citep{boyajian2005}, we
thought the dynamical interaction scenario unlikely based on the low mass
ratios, $q < 0.3$, of HD 14633 and HD 15137.  There are no known O stars
in close, spectroscopic binaries that contain main sequence or pre-main sequence companions later than mid B-type stars \citep{pourbaix2004}, and binary O-type stars usually form in
open clusters with $q \sim 1$ (\citealt{clarke1992} and references
therein).  Therefore it is unlikely that these binaries were born with low
$q$ and then ejected.  However, the current mass ratios can be explained
by dynamical interaction scenerios that exchange binary partners (e.g.\
\citealt*{gualandris2004}).  The pairs may have formed in separate
exchanges involving a wide O-type pair and a close B-type pair with a
comparable binding energy.  Such interactions can also explain the high
$e$ of the resulting binaries.  They would retain this eccentricity for
more than $10^{10}$ years since the low mass ratio significantly increases
the timescale of tidal circularization of the orbit \citep{hilditch2001}.

To test the plausibility of this dynamical origin, we simulated the
ionization of a wide O-type binary in three- and four-body scattering
interactions using the Starlab package developed by P.\ Hut, S.\ McMillan,
J.\ Makino, and S.\ Portegies Zwart.  In each simulation, we assumed a 20
$M_\sun$ + 20 $M_\sun$ pair in a circular orbit with a semi-major axis of
1 AU.  The radius of each O star was set to 8 $R_\sun$ to represent a
typical unevolved pair of late O-type stars.  In our three-body
simulations, we used a projectile consisting of a single 5 $M_\sun$ star
with radius 4 $R_\sun$ and velocity of 140 \kms.  
The B-type projectile with this
velocity at infinity would approach the O star binary with a kinetic
energy, $K$, of only 0.13 times the gravitational potential energy of the
binary, making an ionization extremely unlikely.  In 50,000 three-body
simulations in which we allowed the impact parameter of the projectile,
$r$, to vary from 0 to 5 AU, the target binary became unbound in fewer
than 1\% of our trials.  For $r > 1.25$ AU, the chance of ionization 
decreased to nearly zero.  Increasing $K$ by factors of 2 and 3 (to 
achieve incoming velocities of 200 \kms~ and 240 \kms, respectively) 
reduced the number of ionizations further.

A four-body interaction is far more likely to separate an O star binary.  
For these simulations, we used a target pair identical to that described
above, and the projectile consisted of a 5 $M_\sun$ + 5 $M_\sun$ pair of
B-type stars in a circular orbit with a semi-major axis of 0.12 AU.  We
performed 8200 simulations using the same range in $K$ as above, thereby
allowing incoming velocities of $100-170$ \kms~for the projectile binary,
and allowing $r$ to vary from 0 to 5 AU.  The O binary was ionized in 11\%
of our trials.  We observe no strong dependence on the ionization rate for
$r < 5$ AU with the lowest $K$, although higher $K$ projectiles produce
fewer ionizations when $r > 3$ AU (they tend to fly past the target
binary, preserving the original configurations).

In most of our four-body trials, Starlab predicted collisions between the
targets and/or projectiles, regardless of whether the target binary became
separated.  Without a robust hydrodynamical simulation, it is difficult to
accurately predict how many four-body interactions would result in a
close, bound pair in a binary-binary exchange rather than a stellar
collision.  Nevertheless, our simulations did predict a small number of
exchanges (10) in the four-body scattering scenario.

\section{Conclusions}

Certainly it is plausible that HD 14633 and HD 15137 were both ejected in
multi-body encounters in the dense open cluster NGC 654.  Simulations of a 
typical four-body interaction suggest that 11\% of encounters between low 
mass (5 $M_\sun$ + 5 $M_\sun$) binaries will ionize an O star pair (20 
$M_\sun$ + 20 $M_\sun$).  However, this rate is too low for such a 
dynamical interaction to be probable, and we cannot conclude that these 
binaries were ejected from NGC 654 through such an interaction.

While we do not presently detect a neutron star in either system, we
cannot rule out that these systems were ejected by supernovae in the close
binary systems.  We recommend performing further X-ray observations deep
enough to detect the intrinsic emission from the stellar wind shocks in
these O stars.  Any additional emission from a quiescent neutron star or
from active accretion onto its surface should be comparable to or stronger
than the intrinsic wind emission, thereby producing a measurable X-ray
excess.  Such observations are necessary to determine whether these
systems contain neutron stars and thus determine conclusively which
mechanism is responsible for producing these runaway binary systems.



\acknowledgments

We thank Charles Bailyn, Doug Gies, and the referee, Simon Portegies Zwart, for their
helpful discussions about this work.  The GBT observations would not be
possible without their friendly and diligent staff, especially David Rose,
Becky Warner, and Shirley Curry.  M.\ V.\ M.\ gratefully acknowledges
travel support from NRAO, and she is supported by an NSF Astronomy and
Astrophysics Postdoctoral Fellowship under award AST-0401460.  This
material is based on work supported by the National Science Foundation
under Grants No.~AST-0205297 and AST-0506573 (D.\ R.\ G.).  Institutional
support has been provided from the GSU College of Arts and Sciences and
from the Research Program Enhancement fund of the Board of Regents of the
University System of Georgia, administered through the GSU Office of the
Vice President for Research.  The National Radio Astronomy Observatory is
a facility of the National Science Foundation operated under cooperative
agreement by Associated Universities, Incorporated.  Some of the data
presented in this paper were obtained from the Multimission Archive at the
Space Telescope Science Institute (MAST).  STScI is operated by the
Association of Universities for Research in Astronomy, Inc., under NASA
contract NAS5-26555.  Support for MAST for non-HST data is provided by the
NASA Office of Space Science via grant NAG5-7584 and by other grants and
contracts.  This research has made use of data obtained from the High
Energy Astrophysics Science Archive Research Center (HEASARC), provided by
NASA's Goddard Space Flight Center.

Facilities: \facility{GBT, IUE(SWP), ROSAT(PSPC)}

\input{ms.bbl}

\clearpage

\clearpage
\input{tab1.tex}
\input{tab2.tex}
\input{tab3.tex}
\input{tab4.tex}

\end{document}

%% file: tab1.tex
\begin{deluxetable}{lccccccc}
\tablewidth{0pt}
\tablecaption{Summary of Results from \citealt{mcswain2007} \label{paper1}}
\tablehead{
\colhead{ } &
\colhead{$P$} &
\colhead{ } &
\colhead{$f(m)$} &
\colhead{$a_1 \sin i$} &
\colhead{$d$} &
\colhead{$T_{\rm eff}$} &
\colhead{ }
\\
\colhead{Star} &
\colhead{(d)} &
\colhead{$e$} &
\colhead{($M_\odot$)} &
\colhead{($R_\odot$)} &
\colhead{(pc)} &
\colhead{(K)} &
\colhead{$\log g$} }
\startdata
HD 14633 & 15.4082       & 0.700    & 0.0040    & 4.13    & 2040 & 35100 & 3.95  \\
HD 15137 & 30.35\phn\phn & 0.48\phn & 0.004\phn & 6.7\phn & 2420 & 29700 & 3.50 
\enddata
\end{deluxetable}

%% file: tab2.tex
\begin{deluxetable}{lcccccc}
\tablewidth{0pt}
\tablecaption{Summary of Green Bank Telescope Observations\label{gbt}}
\tablehead{
\colhead{ } &
\colhead{ } &
\colhead{MJD $-$} &
\colhead{Orbital} &
\colhead{Exposure} &
\colhead{$S_{\rm lim}$} &
\colhead{$L_{\rm lim}$} \\
\colhead{Star} &
\colhead{Receiver} &
\colhead{2450000} &
\colhead{Phase} &
\colhead{time (s)} & 
\colhead{($\mu$Jy)} &
\colhead{(mJy~kpc$^2$)} }
\startdata
HD 14633  &  S-band   &  3909.4194  &  0.381  &  6300  &  \phn 15  &  0.062     \\
          &  820 MHz  &  3909.5085  &  0.387  &  6300  &      100  &  0.42 \phn \\
\\
HD 15137  &  S-band   &  3897.4847  &  0.482  &  6300  &  \phn 15  &  0.088     \\
          &  820 MHz  &  3897.5753  &  0.485  &  6300  &  \phn 65  &  0.38 \phn \\
\enddata
\end{deluxetable}

%% file: tab3.tex
\begin{deluxetable}{lcccc}
\tablewidth{0pt}
\tablecaption{Terminal Velocity Measurements\label{termvel}}
\tablehead{
\colhead{ } &
\colhead{\ion{N}{5} $\lambda$1238.821} &
\colhead{\ion{Si}{4} $\lambda$1393.76} &
\colhead{\ion{C}{4} $\lambda$1548.202} &
\colhead{Adopted $V_\infty$} \\
\colhead{Star} &
\colhead{$V_\infty$ (\kms)} &
\colhead{$V_\infty$ (\kms)} &
\colhead{$V_\infty$ (\kms)} &
\colhead{(\kms)} }
\startdata
HD 14633   \dotfill  &  1677    &  \nodata &  \nodata &  $1677 \pm 10$  \\
HD 15137   \dotfill  &  1660    &  1710    &  1690    &  $1690 \pm 25$  \\
\enddata
\end{deluxetable}

%% file: tab4.tex
\begin{deluxetable}{lcc}
\tablewidth{0pt}
\tablecaption{Observed X-ray Flux Limits\label{xraylimits}}
\tablehead{
\colhead{ } &
\colhead{HD 14633} &
\colhead{HD 15137} }
\startdata
Value of nH (cm$^{-2}$)                                            &  $6.76 \times 10^{20}$   &  $2.22 \times 10^{21}$   \\
Upper limit of count rate (cts~s$^{-1}$)                           &  0.00587                 &  0.0129                  \\
Energy band of \textit{ROSAT}/PSPC observation (keV)               &  $0.24-2.0$              &  $0.1-2.4$               \\
\\
Assumed $\Gamma$                                                   &  2.0                     &  2.0                     \\
Unabsorbed $F_X$, $0.5-10$ keV band (erg~cm$^{-2}$~s$^{-1}$)       &  $1.60 \times 10^{-13}$  &  $5.62 \times 10^{-13}$  \\
\\
Assumed $\Gamma$                                                   &  1.6                     &  1.6                     \\
Unabsorbed $F_X$, $0.5-10$ keV band (erg~cm$^{-2}$~s$^{-1}$)       &  $2.39 \times 10^{-13}$  &  $7.85 \times 10^{-13}$  \\
\enddata
\end{deluxetable}